\documentclass[prl,twocolumn,showpacs]{revtex4} 
\usepackage{hyperref} 
\usepackage[dvips]{graphicx,color} 
\begin{document} 
\preprint{APS/123-QED} 
\title{Fluctuation induces evolutionary branching in a 
modeled microbial ecosystem} 
\author{Masashi Tachikawa} 
\email{mtach@complex.c.u-tokyo.ac.jp} 
\affiliation{ 
ERATO Complex Systems Biology Project,  
JST, 3-8-1, Komaba, Meguro-ku, Tokyo 153-8902, Japan. 
} 
\date{\today}

\begin{abstract} 
The impact of environmental fluctuation on species diversity is studied with a model of the evolutionary ecology of microorganisms. We show that environmental fluctuation induces evolutionary branching and assures the consequential coexistence of multiple species. Pairwise invasibility analysis is applied to illustrate the speciation process. We also discuss how fluctuation affects species diversity. 
\end{abstract} 
 
\pacs{ 
87.23.Kg, 
89.75.Fb  
} 
 
\maketitle

Fluctuation is ubiquitous in nature.  
Biological systems are always exposed to environmental fluctuations; 
hence, the systems have evolved 
under these fluctuations \cite{Yan07,Paul00,Kan07}.  
Ecological systems are no exception \cite{GRW02,LK01}.  
Diversity, which is one of the most essential properties in ecology, has
also evolved in the presence of fluctuations.  
The association between diversity and fluctuation has been discussed, and concepts such as intermediate disturbance hypothesis \cite{Con78} have been proposed.  
However, there are few discussions about the mechanisms by which fluctuation is associated with diversity.

In this paper, we study a model of the evolutionary ecology of microorganisms in the presence of a temporal fluctuation.  
We consider competition for a single resource, one of the simplest but most thoroughly investigated ecological situations \cite{SW95,Gro97}. 
In this case it is often believed that only one species can survive by following the competitive exclusion principle \cite{Gau34}. 
However, here we demonstrate that temporal fluctuation in the environment induces evolutionary branching and the stable coexistence of multiple species. 
The results and analysis of the evolutionary dynamics reveal how fluctuation facilitates species diversity.

We adopted a microbial ecosystem because experimentally based quantitative descriptions of its growth kinetics are available \cite{Mon49,KKK98}. 
The growth rate limited by a resource is described by a saturation function called the Monod function \cite{Mon49},  
\begin{eqnarray}  
\mu(s;\tilde{\mu},k) =\tilde{\mu}\frac{s}{k+s}, 
\end{eqnarray}  
where $s$ denotes the concentration of the limited resource. 
The two parameters, the maximum growth rate $\tilde{\mu}$ and the half-saturation constant $k$, are genetically determined and characterize the strategy for the resource utilization of the genotype.

Moreover, recent experimental developments allow us to observe the evolution and ecological diversity of microorganisms directly \cite{KKK98,WWE02,HK06}.  
The data in \cite{KKK98} indicate the relations between parameters of the Monod function among different genotypes, such as the positive correlation $d\tilde{\mu} /d k > 0$, and the logarithmic relation $\tilde{\mu}\propto \log(k)$. 
The former denotes a trade-off relation: fast growth in rich media (large $\tilde{\mu}$) vs. the ability to grow in a wide range of resource concentrations (small $k$). 
Following their work, we introduced a continuous genetic parameter $g$
in our model which specifies the parameters of the Monod function. We chose  
\begin{eqnarray}  
(\tilde{\mu}_g,k_g) =(g, 10^{(g-1)/0.3}),  
\end{eqnarray}  
with an appropriate scale conversion. 
The mutation indicates the small change in $g$. At the population 
level it is described with the diffusion process in $g$ space.

Now we give the model equation by referring to the chemostat experiments
\cite{SW95}. The individual genotype grows at its own growth rate
$\mu(s;\tilde{\mu}_g,k_g)$ and decreases at the same rate $\gamma$. 
The resource $s$ is supplied with the time-dependent function $c(t)$ and
consumed by all populations in proportion to their growth rates. 
Let $x_g$ be the population density of a genotype. The model is given by a partial differential equation with globally interacting variable $s$,   
\begin{eqnarray}  
\left\{  
\begin{array}{l}  
\displaystyle \dot{x}_g =  
\left\{\mu(s;\tilde{\mu}_g,k_g) - \gamma\right\} x_g  +  
D\frac{d^2}{d g^2}  
x_g, \\ 
\displaystyle \dot{s}=c(t) -   
\int \mu(s;\tilde{\mu}_g,k_g) \cdot x_g dg.  
\end{array}\right.   
\end{eqnarray}   
We choose $D=5\cdot 10^{-7}$ and $\gamma=0.11$ in the following simulations.

As the feeding manner, we choose a continuous supply and additions with periodic pulses  
\begin{eqnarray}  
c(t)=c_0+c_1\sum_n \delta(t-nT).  
\end{eqnarray}  
The periodic additions represent the input fluctuation.  
In this paper we choose $T=5$.
We checked that the results are reproduced over a wide range of the period. 
Another parameter, $\lambda$, is introduced to describe the intensity of the fluctuation.  
The amount of supplied resource is given as $c_0=1-\lambda$ and $c_1=\lambda\cdot T$, where the total amount of supply in a period is fixed.

Since we introduced the continuous genotype space, the population is distributed in the space, and the genotype and species do not hold a one-to-one correspondence.  
We use the term quasi-species \cite{ES79} for a group of the population which has a unimodal distribution in the genotype space, and is represented by the genotype at the peak of the distribution. 
If the mutation rate is sufficiently small, the population can be approximately replaced by the population of the representative genotype.

In the simulations, population densities less than $10^{-8}$ are replaced with zero, which denotes the discreteness of the population.

\begin{figure} 
\includegraphics[width=7cm]{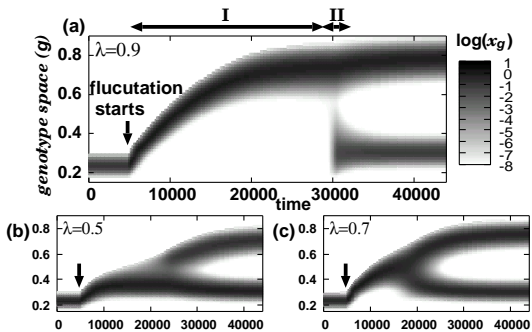} 
\caption{ 
Evolutionary processes as the pattern dynamics of population densities in the genotype spaces.  
The conditions of three different intensities of fluctuation $\lambda=0.5,0.7,$ and $0.9$ are shown.  
The vertical axis denote the genotype space, and the population density normalized in time with period $T=5$ is plotted with gray-scale in a logarithmic scale.  
Initially, static conditions $(c_0,c_1)=(1-\lambda,0.0)$ are set, and stable genotype compositions are provided.  
Fluctuations are introduced at $t=5000$ $(c_0,c_1)=(1-\lambda,5\cdot\lambda)$,  
and evolutionary branching is clearly observed after that. 
} 
\label{fig:1} 
\end{figure}

Fig. \ref{fig:1} shows the evolutionary dynamics in the three different conditions, $\lambda=0.5,0.7,$ and $0.9$. 
In each simulation we initially supply the resource only in a continuous way and the genotype composition of the resident population shows a unimodal distribution in $g$ space; i.e., one quasi-species exists. 
This agrees with the competitive exclusion principle. 
However, introducing fluctuation in the resource supply (the timings are marked with arrows in Fig. \ref{fig:1}), we find dynamic changes of the genotype distributions which lead to evolutionary branching.  
Each of the emerging branches has a unimodal shape and has no connection
with the others.
Therefore, each of them is regarded as an independent quasi-species. 
This implies that the coexistence of multiple quasi-species is attained.

\begin{figure} 
\includegraphics[width=6cm]{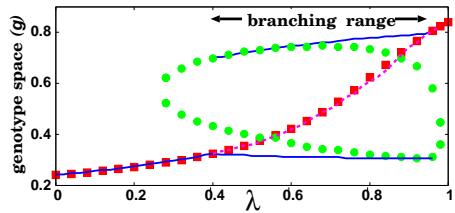} 
\caption{ 
(color online)  
The dependences of characteristic genotypes on the intensity of fluctuation $\lambda$.  
Data from the evolutionary simulations are plotted with lines. 
The representative genotypes at the end of the first phase are plotted with dashed lines, and the representative genotypes at the final state are plotted with continuous lines.  
Characteristic values of PIPs (discussed below), $g_c$ and $g_{\rm ex}$, are plotted with red squares and green circles, respectively. 
$g_c$ values show good correspondences with genotypes at the end of first phase, $g_{\rm ex}$ values are related with genotypes at coexisting states. 
} 
\label{fig:2} 
\end{figure}

When the periodic supply is introduced, the genotype distribution changes through two phases, gradual evolution and branching (indicated by I and II in Fig. \ref{fig:1}-a).  
In the first phase, the distribution moves gradually to higher $g$ in the genotype space, keeping a unimodal shape, which indicates the gradual evolution of the quasi-species. 
The branching starts after the movement stops.

The final states of the three figures in Fig. \ref{fig:1} have similar genotype compositions. Therefore, the states with two quasi-species are robust against the intensity of fluctuation $\lambda$.  
However, features of evolutionary branching dynamics are different.  
The differences are mainly characterized by the states at the onset of branching  
(the states at the end of the first phase). 
In Fig. \ref{fig:2}, we plot lines of the genotype representing the quasi-species  
at the end of the first phase and those of the genotype(s) representing the quasi-species at the final states against $\lambda$.  
Branching was observed in the range $\lambda\simeq 0.40\sim 0.96$.  
At the no-branching conditions, only the first phase appears, and the two lines coincide.

\begin{figure} 
\includegraphics[width=8.5cm]{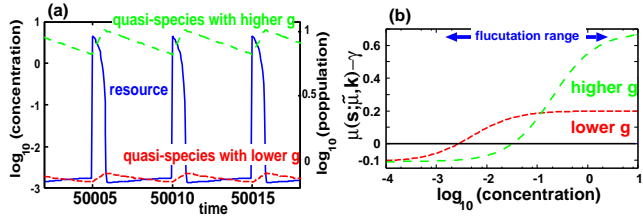} 
\caption{ 
(color online)  
(a) Oscillation of the resource concentration (solid line) and populations of the quasi-species (dashed lines) for $\lambda=0.9$.  
Data are taken after the branching dynamics are completed.  
(b) The Monod function for genotypes representing these quasi-species.  
The fluctuation range of resource concentration is indicated.  
In this range, the faster-growing species change as a consequence of the trade-off relation. 
} 
\label{fig:3} 
\end{figure}

The oscillatory time series of the system after it reaches the final stationary state for $\lambda=0.9$, and the Monod function for the resident quasi-species are shown in Fig. \ref{fig:3}.   
At this state, the two quasi-species have different strategies based on the trade-off relation.  
The difference in the available ranges of resource concentration represents the difference of strategies: a strategy adapted for high concentration and a strategy adapted for a wide range of concentration. 
Therefore, their coexistence can be regarded as a separation of niches in the resource concentration space. 
The temporal fluctuation enables them to utilize the full extent of the concentration space.

In the following, we investigate the mechanism of evolutionary branching dynamics and discuss how fluctuation changes the utility of environmental states.

We start the analysis with the nonfluctuating condition, which corresponds to the situations before the periodic fluctuations are added in the three simulations in Fig. \ref{fig:1}. 
In this condition, one quasi-species is resident.

Suppose a system containing one genotype $g=\alpha$ without mutation. The resource concentration comes to an equilibrium $\overline{s}_{\alpha}$ which holds   
$\mu(\overline{s}_{\alpha};\tilde{\mu}_{\alpha},k_{\alpha})-\gamma=0$. 
Note that $\overline{s}_{g}$ depends only on the traits of the genotype and is independent from the amount of supply $c_0$. 
Thus $\overline{s}_{g}$ is the value characteristic to the genotype. 
If another genotype $g=\beta$ with a smaller equilibrium $(\overline{s}_{\beta}<\overline{s}_{\alpha})$ is introduced into the state,  
the population of $\beta$ grows ($\mu(\overline{s}_{\alpha};\tilde{\mu}_{\beta},k_{\beta})-\gamma>0$).  
This raises the total consumption of the resource, and the resource
concentration decreases.
Then the genotype $\alpha$ decreases and the system comes to a new equilibrium state with the population of genotype $\beta$ and the resource concentration $\overline{s}_{\beta}$.  
The monotonic increase of the Monod function guarantees these processes.

Therefore, the relation between genotypes is determined by the values of $\overline{s}_{g}$, and the genotype $g_c$ which has the minimum value is the fittest genotype. 
The resident genotypes are replaced with genotypes with lower equilibria
sequentially until genotype $g_c$ is reached, 
and the resource concentration comes to $\overline{s}_{g_c}$. 
In our model $g_c\simeq 0.24$ corresponds to the representative genotype of  
quasi-species at the initial static conditions in simulations. 

When fluctuation in the resource supply is introduced, the above simple discussion is not directly applicable. 
This is because the growth or decay of populations depends on
oscillation patterns 
(as an example of the oscillation profile, see Fig. \ref{fig:3}-a). 
Such a situation is in contrast to the above condition, where only the static value of $s$ determines the growth rate. 
To analyze the fluctuation conditions we introduce the pairwise-invasibility plot (PIP), 
which gives the relation between genotypes and enables us to describe the evolutionary dynamics with the process of invasion and annihilation.   
The discussions are based on Geritz {\it et al.} \cite{GM97}.

\begin{figure} 
\includegraphics[width=6cm]{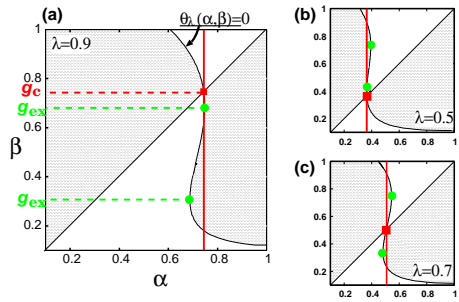}\\ 
\caption{ 
(color online)  
Pairwise invasibility plot for $\lambda=0.5,0.7$, and $0.9$ (corresponding to Fig. \ref{fig:1}).  
$g_c$ and $g_{\rm ex}$ are plotted with red squares and green squares, respectively.  
Vertical lines on $g_c$ (red solid lines) pass through the shaded regions, which denotes  
genotypes invasible to $g_c$. 
} 
\label{fig:4} 
\end{figure}

PIP is constructed with the following procedure.  
Suppose a system with one genotype $g=\alpha$, which is given by  
\begin{eqnarray}  
\left\{  
\begin{array}{l}  
\displaystyle \dot{x}_{\alpha} =  
\left\{\mu(s;\tilde{\mu}_{\alpha},k_{\alpha})  
- \gamma\right\} x_{\alpha},\\  
\displaystyle \dot{s}=c(t) -  
\mu(s;\tilde{\mu}_{\alpha},k_{\alpha}) x_{\alpha}.  
\end{array}\right.  
\end{eqnarray}  
Integration gives a periodic oscillation function of resource concentration  
$\overline{s}_{\alpha}(t)$ with period $T$ after some relaxation time.  
If a small population of another genotype $g=\beta$ is introduced into the system, the average growth rate of $\beta$ is calculated by  
\begin{eqnarray}  
\sigma(\alpha,\beta)=\frac{1}{T} \int_0^T \left\{  
\mu(\overline{s}_{\alpha}(t);\tilde{\mu}_{\beta}, 
k_{\beta})-\gamma\right\}dt.  
\end{eqnarray}  
The sign of $\sigma(\alpha,\beta)$ determines the invasibility of $\beta$ into the $\alpha$ population.  
Calculating $\sigma(\alpha,\beta)$ for all pairs of genotypes and shading the areas of positive $\sigma(\alpha,\beta)$ on the $\alpha$-$\beta$ plane, PIP is obtained. 
The PIPs corresponding to Fig. \ref{fig:1} are shown in Fig. \ref{fig:4}.

Here we investigate the structures of the PIPs and use them to illustrate evolutionary dynamics. 
The PIPs are constructed with two boundary lines: one is a diagonal line $\alpha=\beta$ \footnote{The diagonal line in PIP indicates the invasibility to oneself  
and is always the boundary in PIP ($\sigma_{\lambda}(g,g)=0$).},  and the other is a curved line  
\begin{eqnarray}  
\theta_{\lambda}(\alpha,\beta)=0, \label{eq:7} 
\end{eqnarray}
the shape of which determines the characteristics of PIP. 
First, we refer to the genotype at the intersection between lines as $g_c$ \footnote{Without fluctuation ($\lambda=0$), the intersection coincides with the fittest genotype $g_c$ discussed above.}, which is important because of the following singular property.  
Any genotype lower than $g_c$ is invaded by genotypes higher than it (area above the diagonal line is shaded) and any genotype higher than $g_c$ is invaded by genotypes  
lower than it (area below the diagonal line is shaded).  
This indicates that, as part of the evolutionary process, a resident genotype is replaced by invading genotypes closer to $g_c$ one after another, and the genotype converges to $g_c$.  
Therefore $g_c$ is called the convergent stable genotype \cite{GM97}.  
The convergent process corresponds to the gradual evolution of the quasi-species at the  
first phase in simulations.  
It stops when the representative genotype agrees with $g_c $. The agreement between them is shown in Fig. \ref{fig:2}.  

In Fig. \ref{fig:4}, the vertical line on $g_c$ passes through the shaded regions. 
This means that there are genotypes invasible for the population of $g_c$, and this invasibility promotes evolutionary branching. 
At the end of the first phase the population distributes around $g_c$ in the genotype space. 
If the tail of the distribution covers the region of invasible genotypes, the population of these genotypes grow and form another branch. 
Therefore, evolutionary branching succeeds as the second phase.

The features of branching dynamics are characterized by the relation between $g_c$ and genotypes invasible to $g_c$. 
In PIP for $\lambda=0.5$ genotypes higher than $g_c$ are invasible to $g_c$, 
both higher and lower genotypes are invasible to $g_c$ for $\lambda=0.7$, and lower genotypes are invasible to $g_c$ for $\lambda=0.9$. 
Correspondingly in Fig. \ref{fig:1}, the upper branch is born from the lower one for $\lambda=0.5$, the unimodal shape separates symmetrically for $\lambda=0.7$,  
and the lower branch is born from the higher one for $\lambda=0.9$.

In addition, the resident quasi-species at the final state are related with genotypes $g_{\rm ex}$, which give the local maximum or minimum for $\alpha$ in eq. (\ref{eq:7}),  
\begin{eqnarray}   
\left.  \frac{\partial \theta_{\lambda}}{\partial \beta}  
\biggl/  
\frac{\partial \theta_{\lambda}}{\partial \alpha}  
\right|_{\beta=g_{\rm ex}}=0. 
\end{eqnarray} 
These genotypes can be the locally fittest genotype in the sense that they can invade into the maximum ranges of genotypes (shaded regions bounded by eq. (\ref{eq:7}) have local maximum widths at the points).  
We plot $g_{\rm ex}$ in Fig. \ref{fig:2}.  
In the branching range of $\lambda$, the representative genotypes for the final states stay in the vicinities of $g_{\rm ex}$. 
In particular, at the onset of the branching range, one quasi-species coincides with $g_c$ and the other coincides with $g_{\rm ex}$.  
This is because at the onset, the curved boundary line $\theta_{\lambda}$ 
is tangent to the vertical line on $g_c$ at $\beta=g_{\rm ex}$, and an additional quasi-species arises at the tangent point.

In summary, evolutionary branching dynamics induced by environmental fluctuation is reported in a model of a microbial ecosystem competing for a single resource. 
A pairwise-invasibility plot, which was introduced in \cite{GM97} for a static environment with multiple parameters, is extended to a fluctuating environment and  
is applied here to illustrate evolutionary dynamics.  
 
Previous studies \cite{LK01,SW95} have reported the coexistence of two species with given parameters in the presence of fluctuation.  
However, the occurrence of the branching and evolutionary stability of the coexisting state have remained open questions. Here, we give clear demonstrations of them.  

Although we use a model of a microbial ecosystem, 
the branching mechanism we found 
does not depend on the detail of the model. 
Therefore, the phenomena must be found in more general classes of 
ecosystems.

MacArthur and Levins \cite{ML64} indicated that the number of independently adjustable environmental parameters corresponds to the maximum number of coexisting species as an extended and refined version of the competitive exclusion principle.  
Applying it to our system for the static case, resource concentration is only the adjustable parameter, and it leads to the existence of only one quasi-species.  
However, the stable co-existence of two quasi-species is seen in cases where fluctuation is present.  
This suggests that the number of adjustable parameters is increased by introducing environmental fluctuation.  
When the system is always in an oscillatory state, whether one can grow or not depends on the comprehensive details of the oscillation profile of the parameter.  
In other words, the whole oscillation profile is the environmental state to be adjusted.  
The variety of the profile can be described with several dimensional
parameters which forms a subspace of a functional space.
In this way, introducing fluctuation expands the dimensions of
adjustable environmental parameters, and it enables branching and the coexistence of genotypes with different strategies. 
The diversity of oscillation profiles is not restricted to 
two-dimensional parameter space.
Therefore, by the principle, 
the coexistence of more than two species is possible. 
Although oscillation is introduced as an input fluctuation in this work,
the self-sustained oscillation in population dynamics, if one supposes a
system of generating it, can also induce evolutionary branching based on
the same mechanism.

\begin{acknowledgments} 
The author is grateful to K. Kaneko, S. Ishihara, K. Fujimoto, and M. Inoue for their helpful suggestions. 
\end{acknowledgments}

\end{document}